\documentclass[12pt]{article}
\usepackage{amsfonts}
\usepackage{mathrsfs}
\usepackage{amsthm}
\usepackage{multirow}
\usepackage{bbding}
\usepackage{amssymb}
\usepackage{amsmath}
\usepackage{graphicx,color}
\newcommand{\bea}{\begin{eqnarray}}
\newcommand{\eea}{\end{eqnarray}}
\newcommand{\Bea}{\begin{eqnarray*}}
\newcommand{\Eea}{\end{eqnarray*}}
\newcommand{\ba}{\begin{array}}
\newcommand{\ea}{\end{array}}
\newcommand{\bt}{\begin{tabular}}
\newcommand{\et}{\end{tabular}}
\newcommand{\btb}{\begin{table}}
\newcommand{\etb}{\end{table}}
\newcommand{\bc}{\begin{center}}
\newcommand{\ec}{\end{center}}
\newcommand{\beq}{\begin{equation}}
\newcommand{\eeq}{\end{equation}}

\makeatletter

\newcommand{\Rmnum}[1]{\expandafter\@slowromancap\romannumeral #1@}
\makeatother

\begin{document}

\title{ Inference
 for  biased models: a quasi-instrumental variable approach  }
\author{
Lu Lin,  Lixing Zhu\footnote{The corresponding author:
lzhu@hkbu.edu.hk. Lu Lin is a professor of Shandong University Qilu Securities Institute for Financial Studies and the School of Mathematics
at Shandong University, Jinan, China. His research was supported by NNSF project (11171188 and 11231005) of China, Mathematical
Finance-Backward Stochastic Analysis and Computations in Financial
Risk Control of China (11221061), NSF and SRRF projects (ZR2010AZ001
and BS2011SF006) of Shandong Province of China and K C Wong-HKBU
Fellowship Programme for Mainland China Scholars 2010-11. Lixing Zhu
is a Chair professor of Department of Mathematics at Hong Kong
Baptist University, Hong Kong, China. He was supported by a grant
from the University Grants Council of Hong Kong, Hong Kong, China.
Yujie Gai is an assistant professor  at
Central University of Finance and Economics, Beijing, China. Her research was supported by
NNSF project (11201499) of China. The
first two authors are in charge of the methodology development and
material organization.}\, \,
  and Yujie Gai}
\date{}
\maketitle

\vspace{-4ex}

\begin{abstract}

For linear regression models who are not exactly sparse in the sense that the coefficients of the insignificant variables are not exactly zero, the working models obtained by a variable selection are often biased. Even in sparse cases, after a variable selection, when some significant variables are missing, the working models are  biased as well. Thus, under such situations, root-$n$ consistent estimation and accurate prediction could not be expected. In this paper, a novel remodelling method is proposed to produce an unbiased model when quasi-instrumental variables are introduced. The root-$n$ estimation consistency and the asymptotic
normality can be achieved, and the
prediction accuracy can be promoted as well.  The performance of the new method is examined
through simulation studies.

\


%

\vspace{1ex}

{\it Running head.} Non-sparse regressions.

\end{abstract}

\baselineskip=20pt

\newpage

\setcounter{equation}{0}
\section{ Introduction}

The research described here is particularly motivated by variable selection, but not limited to this area.  Consider the  linear regression model:
\begin{equation}\label{1.1}Y=\beta^{\tau}X+\varepsilon. \end{equation} The model is
seen as the full model that contains all
possibly relevant predictors $X^{(1)}, \cdots,$ $ X^{(p)}$ in the
predictor vector $X$, implying $E(\varepsilon |X)=0$. Here the dimension $p$ of $X$
is large and
even larger than the size $n$ of an available sample.
As is well known,  the root-$n$ estimation consistency
and the asymptotic normality play very important role for  further data
analyses such as confidence region and prediction interval
constructions. However, when there are too many predictors, that is,
when $p$ is too large, even $p>n$, the root-$n$ estimation
consistency is often impossible, and prediction could  be inaccurate.
Variable selection is necessary for us to use those ``kept predictors"
in a working model such that the data analyses can go forward. As such,
variable selection is usually used to remove other  predictors out from the
full model. We will call them the ``removed predictors". Without loss of generality,
suppose that the first
$q$ predictors, $X^{(1)}, \cdots, X^{(q)}$, are kept in a working model whereas the
last $p-q$ predictors are removed from the full model (\ref{1.1}) via a variable
selection approach. $X$ is
then partitioned to be $X=(Z^{\tau},U^{\tau})^{\tau}$, where
$Z=(X^{(1)},\cdots,X^{(q)})^\tau$ and
$U=(X^{(q+1)},\cdots,X^{(p)})^\tau$. Correspondingly, $\beta$ is
partitioned as $\beta=(\theta^{\tau},\gamma^{\tau})^{\tau}$. Thus,
the working model is  the following post-selection model:
\begin{equation}\label{1.2}Y=\theta^{\tau}Z+\eta,\end{equation} where $\eta$ is a new
error term.  Here
working model means that after selection, the model is used to
describe the relationship between the predictor vector $Z$ and the
response $Y$. Note that in this modeling, the error $\eta$ can be
rewritten as $\eta=\gamma^\tau U+\varepsilon$ which in effect
contains all the removed  predictors of $U$. When the full model is sparse such that
$\gamma\equiv 0$, there are a great number of research works available in the literature
to obtain root-$n$ consistent estimation and oracle property, see e.g. the LASSO and the
adaptive LASSO (Tibshirani 1996; Zou
2006), the SCAD (Fan and Li 2001; Fan and Peng 2004), the Dantzig
selector (Cand\'es and Tao 2007) and its relevant development, and the
MCP (Zhang 2010).

{{
However, when $Z$ is
correlated with some components of $U$, the following  may occur:
\begin{equation}\label{1.3}E(\eta|Z)\neq 0.\end{equation} It implies that model (\ref{1.2}) could be biased.  This problem could appear in two scenarios. First, in most of cases in practice, the full model cannot be exactly sparse. Thus, many ``insignificant predictors" with ``small but not nonzero" coefficients are removed in selection procedure. However, as Leeb and P\"otscher
(2005) showed that even when some coefficients are of order $n^{-1/2}$,
the conventional model selection consistency may go wrong.
 Zhang and Huang (2008) also considered the model when there are many coefficients that are small. Under a condition that controls the magnitudes of the small coefficients, several properties about estimation consistency are discussed, but the root-$ n$ consistency cannot be ensured either. It is easy to verify that when the coefficients are of the order $n^{-1/2}$, their condition is not satisfied. In other words, in their setting, even when the coefficients that are associated with ``insignificant predictors" would have a smaller rate  than $n^{-1/2}$, achieving the root-$n$ consistency is still a challenge.  Further, a more practical issue is that even when the model is sparse, any variable selection method would miss some ``significant predictors" and may cause the working model to be biased. Because of the model bias, it is difficult for the
confidence region construction for the
coefficients that are associated with the kept predictors in the working model. Also, the prediction accuracy may be deteriorated.
These observations motivate us to consider how to consistently estimate $\theta$ when the coefficients that are associated with the removed predictors are of slower rate than $n^{-1/2}$ and $E(\eta|Z)$ is not asymptotically negligible. To the best of our knowledge, none of  existing results handle bias correction in the literature.}

Of primary interest in the present paper is to correct the model bias for the working model.  To correct the bias, our idea is to introduce quasi-instrumental variables for bias correction. Note that  those removed predictors  may contain the information about the response and the kept predictors in the working model.
We then determine some quasi-instrumental
variables as functions of these removed predictors. It will be seen that
the use of quasi-instrumental variables makes the re-constructed
model to be an unbiased partially linear model, which is different in
structure from the full model. This
partially linear modeling is of course different from the classical
partially linear modeling in which the predictors in nonparametric
component are given.

{
We should emphasize the following three points for our study.
\begin{itemize}
\item First, if  the number of kept predictors is a fixed value, our partially linear modeling can be directly applied. In other words, practically, our method is always feasible.
\item Second, for those post-selection models, the number of kept predictors is often random. Thus we should assume the model identifiability for us to do further statistical analyses. As an example, we introduce some regularity conditions under which the working model selected by the Dantzig selector (DS) can be identifiable as the sample size goes to infinity. The same idea can be applied to other variable selection methods such as the LASSO, the SCAD or the MCP with different conditions accordingly. It is worth pointing out that the condition allows small coefficients to tend to zero at a rate of order $n^{(c-1)/2}$ for a constant $0<c<1$.  This shows the importance of bias correction and remodeling suggested in the present paper because as was commented on the results from Leeb and P\"otscher
(2005)  and Zhang and Huang (2008), under this setting, existing methods cannot ensure the root-$n$ estimation consistency and asymptotic normality.
\item Third, compared with the commonly used estimations for linear models, our method may be more computational intensive. However, to avoid the risk of possible
unreliable modeling and analysis caused by bias, the cost is worthwhile to pay.
\end{itemize}
}
This work may be a first attempt to achieve the root-$n$ estimation
consistency and then reliable further statistical analyses, there
are several issues that deserve further investigations. The current version of this paper is an updated version of an early manuscript by Lin, Zhu and Gai (2010).

The rest of the paper is organized as follows. In Section~2, when any conventional selection procedures such as the Dantzig selector is used, identifiability conditions are presented, which is particularly designed for the working model when the model size is random, a bias-corrected working modelling
is proposed and a method about
constructing quasi-instrumental variables is suggested. In Section~3, the estimation and
prediction procedures for the new working model are given and the
asymptotic properties of the resulting estimation are
obtained. In Section~4, an method about how to construct low
dimensional nonparametric function is introduced and an approximate
algorithm for constructing quasi-instrumental variables is proposed
for the case where the dimension of the related nonparametric
estimation is relatively large. Simulation studies are presented in
Section~5 to examine the performance of the new approach when
compared with the classical DS and other methods. The technical
proofs for the theoretical results are provided in the online
supplement to this article.

\setcounter{equation}{0}
\section{Identifiability and remodeling}

\subsection{Identifiability: practical and theoretical consideration}
{
As was discussed  in Section~1, when $\theta$ in the working model~(\ref{1.2}) is not random, the identifiability is not necessary and our remodeling to remove bias is feasible.  However, from the theoretical consideration,
existing variable selection approach usually produces a random number of kept predictors in the working model. All existing methods require this random number converges to a non-random constant in a probability sense. Thus, this identifiability issue in our study also asks that the
selected coefficient vector $\theta$  tends to a non-random vector in a probability sense.   This is for theoretical development and the details are presented below.}
\subsection{Model identifiability}

Let
$|J|$ be the number of elements in an index set $J\subset
\{1,2,\cdots,p\}$ and $\bar J$ be the complement of $J$ in the set
$\{1,2,\cdots,p\}$. For a $p$-dimensional vector
$\delta=(\delta_1, \cdots, \delta_p)^{\tau}$, denote by $\delta_J=(\delta_j)_{j\in J}$ a
sub-vector whose entries are those of $\delta$ indexed by $J$. For the full model (\ref{1.1}),
let $I$ be an index set of ``significant regression coefficients" with size $|I|=q$,
and $\bar I$ be its complement in $\{1,\cdots, p \}$ with size $|\bar I|=p-q$. It is worth pointing out that the definition of ``significant regression coefficients" here is different from the classical one because
if the full model is not exactly sparse, the vector of ``insignificant coefficients" with indices in $\bar I$ may not be a zero vector.  The detail of the definition will be given in Proposition~2.1 later. Thus, as a generic methodology, we will  handle biased working
model. It is clear that $I$ is usually unknown, and therefore an estimate
$\tilde I$ of $I$ is required to identify $I$. We need to assume that
$\tilde I$ converges to $I$ in a certain sense such that the corresponding working model
(\ref{1.2}) is seen as to be identifiable.

Afterwards, a bias correction can be performed. Clearly, a natural
question is how to get  an estimate $\tilde I$ in practice. 
When the full model (\ref{1.1}) is exactly sparse, ``significant
regression coefficients" are those nonzero coefficients and thus they are well defined.
When the full model is not exactly sparse,  we must distinguish between ``significant regression
coefficients" and ``insignificant regression coefficients" such that the index set $I$ can
be well defined and consistently estimated. For a variable selection procedure, $I$ may be
a set $\{1\leq j\leq p:|\beta_j|\geq
\zeta_n\} $ for a positive value $\zeta_n$ that may depend on the sample size $n$. Later,
we will present an example using the Dantzig selector
to identify such a set.
Denote by $\tilde \beta_j$ an estimate of the $j$-th component $\beta_j$ of
the regression coefficient vector $\beta$ in the full model (\ref{1.1}) via a selection method such as the LASSO or the SCAD or the Dantzig selector. An estimate  of $I$ can be defined as $$\tilde I_{\tau_n}=\{1\leq j\leq p:|\tilde\beta_j|\geq
\tau_n\},$$ where $\tau_n>0$ is a predetermined threshold that will depend on which variable selection approach is applied in the variable selection procedure.
The required
identifiability condition  can be written as:
\begin{itemize} \item[\it C0.] For the full model (\ref{1.1}), if  there exists a threshold $\tau_n>0$ satisfying
$$P(\tilde I_{\tau_n}=I)\rightarrow 1\ \mbox{ as } \ n \rightarrow \infty,$$
the corresponding working model (\ref{1.2}) is then called to be identifiable.
\end{itemize}  Under this condition,  $\tilde I_{\tau_n}=I$ holds
with probability tending to 1. Thus,  the working model (\ref{1.2}) can
be non-random  in an asymptotic sense. {
In the following, we use the Dantzig selector as an example to  explain the identifiability condition when the model may not be exactly sparse.}

{\it Identifiability condition for the Dantzig selector.} Denote $\sigma^2=E(\varepsilon^2|X)$
and let $\lambda_p$ be the tuning parameter
used for constructing the Dantzig selector estimation.
We first suppose without loss of generality that all the diagonal elements of
the matrix ${\bf X}^{\tau}{\bf X}/n$
are equal to 1. The key idea about the identifiability conditions
for the Dantzig selector is to clearly distinguish between the kept predictors and removed
predictors whether the full model is exactly sparse or not. The conditions and result
are stated in the following proposition.

{\bf Proposition 2.1} {\it For the full model (1.1), assume $p=\exp(n^c)$ for a constant
$0<c<1$ and the following conditions hold:
 \begin{eqnarray}\label{2.1}\left\{\begin{array}{ll}\min_{\delta_{I}\neq 0, \|\delta_{\bar I}\|_{\ell_1}\leq
\|\delta_{I}\|_{\ell_1}+c_2n^{(c-1)/2}}\frac{\|{\bf X}\delta\|_{{\ell}_2}}{\sqrt{n}
\|\delta_{I}\|_{{\ell}_1}}>\sqrt{c_1},\vspace{1ex}\\
\min\limits_{j\in I}|\beta_j|
\geq c_3 n^{(c-1)/2}, \, \, \|\beta_{\bar I}\|_{\ell_1}\leq c_2n^{(c-1)/2},\end{array}\right.
\end{eqnarray} and moreover, the regression coefficient vector $\beta$ is estimated by the Dantzig selector $ \tilde\beta^D$ with $\lambda_p=2\sigma\sqrt{\frac{\log p}{n}}$ and the index set $I$ is estimated by
 $\tilde I_{\tau_n}=\{1\leq j\leq p:|\tilde\beta_j^D|\geq
\tau_n\}$ with $\tau_n=c_4n^{(c-1)/2}$, where positive constants satisfy
 $c_3/2>c_4> \left( 4\sigma/c_1+ 3c_2+\sqrt {8c_2/c_1+8\sigma^2/c_1^2}     \right).$
Then, as
$n \rightarrow \infty$,
$$P(\tilde I_{\tau_n}=I)\rightarrow 1.$$
}

{\bf Remark 1.} {\it {
It is noted that the inequalities in (\ref{2.1}) are in sprit similar to those in Bickel {\it et al} (2009). However, the key difference  is that the minimum is not over all $j$ with $1\le j\le p$ while over a subset $I$ that is often small compared with $\bar I$. This condition weaker than that of Bickel {\it et al} (2009) plays an important role, otherwise,  the $p\times p$ matrix ${\bf X}{\bf X}^T/n$ is required to be positive definite and when $p>n$, this requirement will not be possible. The separation rate is $n^{(c-1)/2}$ for $0<c<1$. When $c$ is large, the corresponding coefficients of the removed predictors can be of the order $n^{(c-1)/2}$. As was commented before, under this condition, any conventional estimation such as the estimation in Zhang and Huang (2010) cannot achieve the root-$n$ consistency.
 }}


{
The proof of the proposition is given in the supplement. It is worth pointing out that other variable selection methods, such as the LASSO, can also be used as examples for this purpose. But the identifiability condition may be different. The detailed conditions deserve a further study. } 

\subsection{Re-modeling}

To correct the bias of the  working model (\ref{1.2}) for constructing
a root-$n$ consistent estimate of the sub-vector
$\theta=(\theta_1,\cdots,\theta_q)^{\tau}$, we establish a new
model with a quasi-instrumental variable introduced by us. The details are described in this
subsection.

As was discussed before, we assume that the $q$-dimensional significant predictor
vector $Z$ is  selected into the working model with a probability
going to one. From now on we suppose that the dimension $q$ is non-random  for simplicity. If $q$ depends on $n$ but is much smaller than $n$, the method proposed below is still valid. We further assume that
$$\Sigma_{Z,Z}=I_q,$$ where $\Sigma_{Z,Z}=Cov(Z)$ and $I_q$ is a $q\times q$ identity matrix. This assumption is common because the dimension $q$ of the covariate $Z$ in the working model (\ref{1.2})
is low and furthermore, a standardization over $Z$ does not affect the theoretical development. Our idea is to select some predictors from the removed predictors to define the quasi-instrumental variable. We give its outline below.

{\it 1). Re-modeling}

 Define
$$\tilde Z=(Z^{\tau},\tilde U^{\tau})^{\tau},$$ where $\tilde
U=(\Sigma_{U^*,U^*}-\Sigma_{U^*,Z}\Sigma_{Z,U^*})^{-1/2}(U^*-\Sigma_{U^*,Z}Z)$,
$\Sigma_{U^*,U^*}=Cov(U^*)$, $\Sigma_{U^*,Z}=Cov(U^*,Z)$ and
$U^*=(U^{(k_1)},\cdots,U^{(k_d)})^\tau$ for some integer $d\ge 1$
when the vectors $U^*$ and $Z$ are not linearly correlated. Here
$U^{(k_1)},\cdots,U^{(k_d)}$ are $d$ random variables selected from $U$.
The detailed selection method and conditions on
$U^{(k_1)},\cdots,U^{(k_d)}$ will be given in Section 4. Thus, we
can see that $\tilde U$ is uncorrelated with $Z$ and its covariance
matrix is an identity matrix. This leads to an identity matrix as
the covariance matrix of $\tilde Z$: $\Sigma_{\tilde Z, \tilde Z
}=I_{q+d}$.
 Let
$\Sigma_{U,\tilde Z}=Cov(U,\tilde Z)$ and denote by $r$ the rank of
matrix $\Sigma_{U,\tilde Z}$. Write
$$ V=A\tilde Z,$$ where $A$ is an $r\times (q+d)$ matrix to be
be specified later. Suppose that the selected $A$ and
$U^{(k_1)},\cdots,U^{(k_d)}$ satisfy
\begin{equation}\label{2.2}E\{(Z-E(Z|V))(Z-E(Z|V))^{\tau}\}>0.\end{equation}
This condition on the matrix can trivially hold because $V$ is a
weighted sum of $Z$ and $U^{(k_1)},\cdots,U^{(k_d)}$. Condition (\ref{2.2}) helps identify the following
model. Further conditions on $A$ will be discussed later.

Denote $g(V)=E(\eta|V)$.  A
bias-corrected version of (\ref{1.2}) is defined as
\begin{equation}\label{2.3} Y=\theta^{\tau}Z+g(V)+\xi(V), \end{equation} where $\xi(V)
=\eta-g(V)$. As described above, we must find a vector $V=A\tilde Z$ such that  $E(\xi(V)|Z,V)=0$. This appended vector
$V$ could be regarded as a quasi-instrumental variable as  it is not exactly an instrumental variable for endogenous variable $Z$ in the classical sense. 
To this end, we must properly select the matrix $A$. We will discuss it below.

If the quasi-instrumental variable $V$ were given,  model (\ref{2.3})
could be regarded as a partially linear model with a linear
component $\theta^{\tau}Z$ and a nonparametric component $g(V)$, and
could be  identifiable because of condition (\ref{2.2}). The model can
then still describe the regression relationship between the
significant predictors $Z$ and the response $Y$ although a
nonparametric function $g(v)$ gets involved.

{\it 2). A brief outline of instrumental variable selection}.

To determine matrix $A$,  we assume  the condition that $(Z,U)$
is elliptically symmetrically distributed. The ellipticity condition
can be slightly weakened to be the following {\it linearity condition}:
\begin{equation}\label{2.4}E(U|C^{\tau}\tilde Z)=E(U)+
\Sigma_{U,\tilde Z}C(C^{\tau}\Sigma_{\tilde Z,\tilde
Z}C)^{-1}C^{\tau}(\tilde Z-E(\tilde Z)) \end{equation} for some given
matrix $C$. The linearity condition has been widely assumed in the
circumstance of high-dimensional models, see Hall and Li (1993).
Note that this condition is not very strong because Hall and Li (1993)
showed that it often holds approximately when the dimension $p$ is
high in the sense that $p\to \infty$ as $n\to \infty$. This is just
the scenario we work on.

Since $\Sigma_{U,\tilde Z}^\tau \Sigma_{U,\tilde Z}$ is a
symmetric matrix, there exists a $(q+d)\times (q+d)$ orthogonal
matrix $Q$ such that $\Sigma_{U,\tilde Z}^\tau \Sigma_{U,\tilde
Z}=Q\Lambda Q^\tau$, where $\Lambda$ is a diagonal matrix, its first
$r$ diagonal elements are positive and the others are equal to zero.
Let $Q=(Q_1,Q_2)$ and $Q_1^\tau=(Q_{11}^\tau,Q_{12}^\tau)$, where
$Q_1$ is a $(q+d)\times r$ matrix, $Q_{11}$ is a $q\times r$ matrix
and $Q_{12}$ is a $d\times r$ matrix. 
The following lemma shows that we can find a matrix $A$ such that
model (\ref{2.3}) is always unbiased.

{\bf Lemma 2.2} {\it \ Suppose that  condition (\ref{2.4}) holds, the
eigenvalues of $Q_{11}^\tau Q_{11}$ are not equal to 1,
$\Sigma_{Z,Z}=I_{q}$, and $U^*$ and $Z$ are not linearly correlated.
When
\begin{equation}\label{2.5}A=Q_1^{\tau},\end{equation} model (\ref{2.3}) is then unbiased in the sense that $E(\xi|Z,V)=0$.}

Note that in many cases an eigenvalue of $Q_{11}^\tau Q_{11}$ may not
exactly equal  1. Thus, the condition on the
eigenvalues of $Q_{11}^\tau Q_{11}$ is mild.  The proof of
Lemma~2.2 is presented in the Supplement. By this lemma, the selected
quasi-instrumental variable is
\begin{equation}\label{2.6}V=Q_1^{\tau}\tilde Z.\end{equation}

We now discuss how to estimate $Q_1$. It is clear that the matrices $\Sigma_{U^*,U^*}$, $\Sigma_{U^*,Z}$ and
$\Sigma_{U,\tilde Z}^\tau \Sigma_{U,\tilde Z}$ are not always given
and then need to be estimated. Thus, $Q_1$ defined in (\ref{2.5}) and $\tilde Z$ in (\ref{2.6}) need to be estimated.  Since the dimensions of the matrices $\Sigma_{U^*,U^*}$ and
$\Sigma_{U^*,Z}$ are much lower than the sample size $n$, and then they can be
consistently estimated, for example by the corresponding sample covariances, and so can $Q_1.$ After
$\Sigma_{U^*,U^*}$ and $\Sigma_{U^*,Z}$ are replaced respectively by
their root-$n$ consistent estimates in the expression of $\tilde
Z$, the estimated value of $\tilde Z$ is obtained. Without any notational confusion, we still use $\tilde Z$ to denote the estimated value of $\tilde Z$. In the matrix $\Sigma_{U,\tilde Z}^\tau
\Sigma_{U,\tilde Z}$, every element is a sum of $p-q$ summands. We will estimate $\frac
1{p-q}\Sigma_{U,\tilde Z}^\tau \Sigma_{U,\tilde Z}$ to have a consistent estimate of $Q$ because of the invariance of    $Q$  for
$\frac 1c\Sigma_{U,\tilde Z}^\tau \Sigma_{U,\tilde Z}$ for any positive value $c>0$.
Without loss of generality,  assume that $Z$ and $U$ have zero mean. Let $U_1,\cdots,U_n$, $Z_1,\cdots,Z_n$ and $Y_1,\cdots,Y_n$ be the samples of $U$, $Z$ and $Y$ respectively. Consequently, we get a sample of $\tilde Z$ as $\tilde Z_1,\cdots,\tilde Z_n$.
Note that the dimension of $\Sigma_{U,\tilde Z}^\tau
\Sigma_{U,\tilde Z}$ is fixed and $E(\tilde Z_1U^\tau_1 U_2{\tilde
Z_2}^\tau)=\Sigma_{U,\tilde Z}^\tau \Sigma_{U,\tilde Z}$. We then
use the following $U$-statistic as an estimate of
$\frac{1}{p-q}\Sigma_{U,\tilde Z}^\tau \Sigma_{U,\tilde Z}$:
$$\frac{1}{p-q}\widehat{\Sigma_{U,\tilde Z}^\tau \Sigma_{U,\tilde Z}}
=\frac{1}{p-q}\frac{2}{n(n-1)}\sum_{1\leq i<j\leq n}\tilde Z_iU^\tau_i
U_j{\tilde Z_j}^\tau.$$ It is clear that the kernel $h(\tilde
Z_1,U_1;\tilde Z_2,U_2)=\frac{1}{p-q}\tilde Z_1U^\tau_1 U_2{\tilde
Z_2}^\tau$ of the above $U$-statistic has mean
$\frac{1}{p-q}\Sigma_{U,\tilde Z}^\tau \Sigma_{U,\tilde Z}$.
Furthermore, the $(i,j)$-th component of the kernel is
$h_{i,j}=\frac{1}{p-q}\sum_{k=1}^{p-q}{\tilde
Z_1}^{(i)}U_1^{(k)}U_2^{(k)}{\tilde Z_2}^{(j)}$, its projection on
$(\tilde Z_2,U_2)$ is
$$E[h_{i,j}|\tilde Z_2,U_2]=\frac{1}{p-q}\sum_{k=1}^{p-q}E[{\tilde Z_1}^{(i)}U_1^{(k)}]
U_2^{(k)}{\tilde Z_2}^{(j)}$$ and the variance of the projection is
\begin{eqnarray*}&&Var\{E[h_{i,j}|\tilde Z_2,U_2]\}\\&&
=\frac{1}{(p-q)^2}\sum_{k=1}^{p-q} E^2[{\tilde
Z_1}^{(i)}U_1^{(k)}]E[U_2^{(k)}{\tilde Z_2}^{(j)}]^2
\\&&+\frac{2}{(p-q)^2}\sum_{k_1<k_2}^{p-q}E[{\tilde Z_1}^{(i)}U_1^{(k_1)}]
E[{\tilde Z_1}^{(i)}U_1^{(k_2)}]E[U_2^{(k_1)} U_2^{(k_2)}({
Z_2}^{(j)})^2].\end{eqnarray*} This shows that if
\begin{equation}\label{2.7}E(U^{(k)}{\tilde Z}^{(j)})^2\leq C,\ \ 1\leq k\leq p-q,1\leq j\leq
q+d,\end{equation} for a constant $C>0$, then
$$\frac{1}{p-q}\widehat{\Sigma_{U,\tilde Z}^\tau \Sigma_{U,\tilde Z}}-
\frac{1}{p-q}{\Sigma_{U,\tilde Z}^\tau \Sigma_{U,\tilde
Z}}=O_p(n^{-1/2}).$$ Moreover, there exists a $(q+d)\times (q+d)$
orthogonal matrix $\hat Q$ satisfying $\widehat{\Sigma_{U,\tilde
Z}^\tau \Sigma_{U,\tilde Z}}= \hat Q\hat \Lambda \hat Q^\tau$, where
$\hat\Lambda$ is a diagonal matrix, its first $\hat r$ diagonal
elements are positive and the others are equal to zero. Thus
$$\widehat{\Sigma_{U,\tilde Z}^\tau \Sigma_{U,\tilde Z}}=
\hat Q\hat \Lambda \hat Q^\tau.$$
Let $\hat Q=(\hat Q_1,\hat Q_2)$ and $\hat Q_1^\tau=(\hat
Q_{11}^\tau,\hat Q_{12}^\tau)$, where $\hat Q_1$ is a $(q+d)\times
\hat r$ matrix, $\hat Q_{11}$ is a $q\times \hat r$ matrix and $\hat
Q_{12}$ is a $d\times \hat r$ matrix. Note that when we use $\frac 1{p-q}\widehat{\Sigma_{U,\tilde Z}^\tau \Sigma_{U,\tilde Z}}$, the decomposition is exactly the same: $\hat Q\frac{1}{p-q}\hat \Lambda \hat Q^\tau$ except for a constant factor $\frac{1}{p-q}.$
Because $\hat
Q\frac{1}{p-q}\hat \Lambda \hat Q^\tau$ is a root-$n$ consistent
estimate of $\frac{1}{p-q}{\Sigma_{U,\tilde Z}^\tau
\Sigma_{U,\tilde Z}}$, it is easy to see that
$$\hat A=\hat Q^\tau_1 $$
is a consistent estimate of $A$. 
By this way, an estimated quasi-instrumental variable
can be defined as
$$\hat V=\hat Q_1^{\tau}\tilde Z.$$

From the above choice of $A$, we can see that $g(v)$ is a
$r$-variate nonparametric function. When $r$ is large, the resulting
nonparametric estimate for $g(v)$ is inefficient. We will introduce
two methods in Section 4 to reduce the dimension of variable $V$.

\setcounter{equation}{0}
\section{ Estimation and asymptotic normality}

\subsection{Estimation} Recall that the bias-corrected model (\ref{2.3}) can be thought of as a
partially linear model and is unbiased when $A=Q_1^\tau$. Then we
can design an estimation procedure as follows. Given $\theta$, if
$A$ is estimated by $\hat A$,
then the nonparametric function $g(v)$ is estimated by
$$\hat g_\theta(v)=\frac{\sum_{k=1}^n(Y_k-\theta^{\tau}
Z_k)L_H(\hat V_k-v)} {\sum_{k=1}^nL_H(\hat V_k-v)},$$ where $\hat
V=\hat A\tilde Z$ and $L_H(\cdot)$ is a $r$-dimensional kernel
function. A simple choice of $L_H(\cdot)$ is a product kernel as
$$L_H(V-v)=\frac{1}{h^{r}}K\Big(\frac{V^{(1)}-v^{(1)}}{h}\Big)\cdots
K\Big(\frac{V^{(r)}-v^{(r)}}{h}\Big),$$ where
$V^{(j)},j=1,\cdots,r$, are the components of $V$, $K(\cdot)$ is an
1-dimensional kernel function and $h$ is the bandwidth depending on
$n$.

With the estimate of $g(v)$, the
bias-corrected model of (\ref{2.3}) can be approximately expressed by the
following model:
$$Y_i\approx \theta^{\tau}Z_i+\hat g_\theta(\hat V_i)+\xi(\hat V_i),$$ equivalently,
\begin{equation}
\label{3.1}\hat Y_i\approx \theta^{\tau}\hat Z_i+\xi(\hat V_i),\end{equation}
where
$$\hat Y_i=Y_i-\frac{\sum_{k=1}^nY_kL_H(\hat V_k-\hat V_i)}
{\sum_{k=1}^nL_H( \hat V_k-\hat V_i)},\ \ \hat
Z_i=Z_i-\frac{\sum_{k=1}^nZ_kL_H(\hat V_k-\hat
V_i)}{\sum_{k=1}^nL_H( \hat V_k- \hat V_i)}.$$ Thus, the working
model in (\ref{3.1}) results in an estimate of $\theta$ as
\begin{equation}
\label{3.2}\hat\theta=\hat S_n^{-1}\frac{1}{n}\sum_{i=1}^n \hat
Z_i\hat Y_i,\end{equation} where $\hat S_n=\frac{1}{n}\sum_{i=1}^n\hat
Z_i\hat Z_i^{\tau}$. Here we assume that the bias-corrected model
(\ref{2.3}) is homoscedastic, that is $Var(\xi(\hat V_i))=\sigma_V^2$ for
all $i=1,\cdots,n$. If the model is heteroscedastic and
$\sigma_i^2(\hat V_i)=Var(\xi(\hat V_i))$ is assumed to be known, we
modify the above estimate to be
$$\hat\theta^*=\hat {S_n^*}^{-1}\frac{1}{n}\sum_{i=1}^n
\frac{1}{\sigma_i^2(\hat V_i)}\hat Z_i\hat Y_i,$$ where $\hat
S_n^*=\frac{1}{n}\sum_{i=1}^n\frac{1}{\sigma_i^2(\hat V_i)}\hat
Z_i\hat Z_i^{\tau}$. When $\sigma_i^2(\hat V_i)$ is unknown, we can
use its consistent estimate to replace it; for details about how to
construct the estimate see for example H\"ardle {\it et al.}
(2000). In the following we only consider the estimate defined in
(\ref{3.2}). Finally, an estimate of $g(v)$ can be defined as $\hat
g_{\hat\theta}(v)$.

Also the bandwidth selection is an issue. Because the remodeled
model above is a partially linear model, the bandwidth selection for
such a model has been sufficiently investigated (see, e.g., Fan and
Huang 2005 and H\"ardle {\it et al.} 2000). Thus the details are
omitted in this paper.

In summary, the algorithm includes the following
three steps:
\begin{itemize}\item[\it S1.] Estimate matrices $\Sigma_{U^*,U^*}$ and
$\Sigma_{U^*,Z}$ by the corresponding sample covariances. Estimate matrix
$\Sigma_{U,\tilde Z}^\tau \Sigma_{U,\tilde Z}$ by the $U$-statistic
given above and decompose the estimated matrix as
$\hat\Sigma_{U,\tilde Z}^\tau \hat\Sigma_{U,\tilde Z}=\hat
Q\hat\Lambda \hat Q^\tau$ and $\hat Q=(\hat Q_1,\hat Q_2)$.
\item[\it S2.] Calculate $\tilde Z$ by replacing $\Sigma_{U^*,U^*}$
and $\Sigma_{U^*,Z}$ in the expression of $\tilde Z$ with their
consistent estimates. Choose the estimate of $A$ as $\hat A=\hat
Q_1^{\tau}$ and construct a quasi-instrumental variable as $\hat
V=\hat Q_1\tilde Z$.
\item[\it S3.]
Construct the estimate of $\theta$ by the formula in
(\ref{3.2}).\end{itemize}

The steps show that the new algorithm is slightly more
complicated to implement than the ones for linear models are. However, to get the root-$n$ estimation consistency and
accurate prediction for non-sparse models, such a cost is worthwhile
to pay.

\subsection{Asymptotic normality}

To study the asymptotic behavior, the following conditions for the
model (\ref{2.3})  are assumed:

\begin{itemize}
\item[\it C1.] The first two derivatives of $g(v)$ and $\xi(v)$ are continuous.
\item[\it C2.] Kernel function
$K(\cdot)$ satisfies $$\int K(u)du=1, \int
u^jK(u)du=0,j=1,\cdots,m-1,0< \int u^mK(u)du<\infty.$$
\item[\it C3.] The bandwidth $h$ is optimally chosen, i.e., $h=O(n^{-1/(2m+r)})$.
\item[\it C4.] The condition (\ref{2.7}) holds.
\end{itemize}

Obviously, Conditions {\it C1-C3} are commonly used for
semiparametric models. Condition {\it C4} is a commonly used moment
condition. Furthermore, $Q_1^{\tau}Q_1=I_r$ and the condition on
eigenvalues of $Q_{11}^\tau Q_{11}$ in Lemma 2.2 implies that
$Q_{12}^\tau Q_{12}$ is invertible. Thus, we have the following
theorem.

{\bf Theorem 3.1} {\it In addition to the conditions in Lemma~2.2,
assume that conditions C1-C4 and (\ref{2.2}) hold and $q$ is fixed. Then, we have that as $n\rightarrow\infty$,
$$\sqrt{n}(\hat\theta-\theta)\stackrel{D}\longrightarrow
N(0,\sigma_V^2(Q_{12}^\tau Q_{12})^{-1}).$$

}

The proof for the theorem is postponed to the Supplement.

{\bf Remark~3.1.} From the theorem, we have the following findings:
\begin{itemize}\item[(1)]
This theorem shows that the new estimate $\hat\theta$ is
$\sqrt{n}$-consistent regardless of the choice of the shrinkage
tuning parameter $\lambda_p$ and thus it is convenient to use in
practice. Furthermore, by the theorem and the commonly used
nonparametric techniques, we can prove that $\hat g_{\hat\theta}(v)$
is also consistent. In effect, we can obtain the strong consistency
and the consistency of the mean squared error under some stronger
conditions. The details are omitted in this paper. Note that these
results can obviously  hold when the model is exactly sparse. Thus,  our method always ensures the root-$n$ estimation consistency
for the coefficients associated with the kept predictors in the
working model. On the other hand, the asymptotic covariance of the
new estimate relies on the choice of the quasi-instrumental
variable $V$. Similarly as existing approaches in the
literature, it is difficult or impossible to choose the optimal
instrumental variable for the asymptotic efficiency. \item[(2)] Note
that the choice of $A$ is not unique because the choice of $\tilde
Z$ is not unique. For a dimension reduction choice of $A$ given in the
next section, we will have some further theoretical results.
\item[(3)] Here we only consider the case where $q$ is fixed for simplicity. It was stated above if $q$ depends on $n$ but is much smaller than $n$, the method proposed above is still valid. In this case the asymptotic normality for the estimator $\Delta\hat\theta$ still holds, where $\Delta$ is a $s\times q$ matrix and $s$ is fixed. \end{itemize}

\subsection{Prediction}

Combining the estimation consistency with the unbiasedness of the
adjusted working model (\ref{2.3}), we obtain an improved prediction as
\begin{equation}
\label{3.3}\hat
Y=\hat\theta^{\tau}Z+\hat g_{\hat\theta}(V)\end{equation} and the
corresponding prediction error is
$$\begin{array}{lll}E(Y-\hat Y)^2&=&E((\hat\theta-\theta)^{\tau}Z)^2+E(\hat
g_{\hat\theta}(V)-g(V))^2+E(\xi^2(
V))\vspace{1ex}\\&&+2E((\hat\theta-\theta)^{\tau}Z(\hat
g_{\hat\theta}(V)-g(V)))+2E((\hat\theta-\theta)^{\tau}Z\xi(
V))\vspace{1ex}\\&&+2E((\hat g_{\hat\theta}(V)-g(V))\xi(
V))\vspace{1ex}\\&=&E(\xi^2( V))+o(1).\end{array}$$ It is of a
smaller prediction error than the one obtained by the classical Dantzig
selector, and interestingly  any high-dimensional
nonparametric estimation is not needed.

In contrast, when we use
the new estimate $\hat\theta$ and the working model (\ref{1.2}), rather
than the adjusted working model (\ref{2.3}), the resulting  prediction is defined as
\begin{equation}
\label{3.4}\hat Y_S=\hat\theta^{\tau}Z+\bar{\hat g}_{\hat\theta},\end{equation}
where $$\bar{\hat g}_{\hat\theta}=\frac{1}{n}\sum_{i=1}^n\hat
g_{\hat\theta}(V_i).$$ We  add $\bar{\hat g}_{\hat\theta}$ in (\ref{3.4})
for prediction because the working model (\ref{1.2}) has a bias $E(g(V))$,
otherwise, the prediction error would be even larger.  In this case,
$\bar{\hat g}_{\hat\theta}$ is free of the predictor $U$ and the
resultant prediction of (\ref{3.4}) only uses the predictor $Z$ in the
working model (\ref{1.2}). This  is different from the prediction (\ref{3.3})
that depends on both the low-dimensional predictor $Z$ and a part of
the high-dimensional predictors $U$. Thus (\ref{3.4}) is a
working model based prediction. The corresponding prediction error
is
$$\begin{array}{lll}E(Y-\hat
Y_S)^2&=&E((\hat\theta-\theta)^{\tau}Z)^2+E(\bar{\hat
g}_{\hat\theta}-g(V))^2+E(\xi^2(
V))\vspace{1ex}\\&&+2E((\hat\theta-\theta)^{\tau}Z(\bar{\hat
g}_{\hat\theta}-g(V)))+2E((\hat\theta-\theta)^{\tau}Z\xi(
V))\vspace{1ex}\\&&+2E((\bar{\hat g}_{\hat\theta}-g(V))\xi(
V))\vspace{1ex}\\&=&E(\xi^2( V))+Var(g(V))+2E(E(g(V))-g(V))\xi(
V))+o(1).\end{array}$$ This error is usually  larger than that of
the prediction (\ref{3.3}). However, we can see that
$$|E(E(g(V))-g(V))\xi(
V))|\leq (Var(g(V))Var(\xi(V)))^{1/2}$$ and usually the values of
both $Var(g(V))$ and $Var(\xi(V))$ are small. Then the prediction error of such a prediction is smaller than that of
\begin{equation}
\label{3.5}\tilde Y_S=\tilde\theta_S^{\tau}Z,\end{equation} which is
obtained by the
working model (\ref{1.2}) and the common LS estimate
$\tilde\theta_S=({\bf Z}^{\tau}{\bf Z})^{-1}{\bf Z}^{\tau}{\bf Y}$.
Precisely, the
corresponding error of $\tilde Y_S$ in (\ref{3.5}) is
$$\begin{array}{lll}E(Y-\tilde Y_S)^2&=&E((\tilde\theta_S-\theta)^{\tau}Z)^2+E(\gamma^{\tau}U)^2+\sigma^2
+2E((\tilde\theta_S-\theta)^{\tau}Z\gamma^{\tau}U).\end{array}$$ Because
$\tilde\theta_S$ does not converge to $\theta$, the values of both
$E((\tilde\theta_S-\theta)^{\tau}Z)^2$ and
$2E((\tilde\theta_S-\theta)^{\tau}Z\gamma^{\tau}U)$ are large and as a result
the prediction error is large as well.

The above results show that in the scope of prediction, the new
method can reduce prediction error under both the adjusted
working model (\ref{2.3}) and the original working model (\ref{1.2}). We will
see that the simulation results in Section~5 coincide with these
conclusions.

\setcounter{equation}{0}
\section{Construction of dimension-reduced variable $V$}

As was mentioned before, when $r$ is large, a $r$-dimensional
nonparametric estimation will be involved, which may lead to
inefficient estimation. Thus a low-dimensional quasi-instrumental
variable $V$ is desired. In this section, we suggest two alternative approaches.

\subsection{Method~1}

From Proposition~2.1 we see that to get a low-dimensional variable $V$, it
is sufficient to construct matrix $\Sigma_{U,\tilde Z}$ such that
its rank $r$ is as small as possible. Assume with no loss of generality that
$\|\mbox{corr}(U^{(1)}Z)\|_{\ell_1}\geq
\|\mbox{corr}(U^{(2)}Z)\|_{\ell_1}\geq \cdots \geq
\|\mbox{corr}(U^{(p-q)}Z)\|_{\ell_1}$.
We then choose $U^*$ as $U^*=(U^{( 1)},\cdots,U^{( d)})^\tau$. Together with  the definition of $\tilde U$ given in
Subsection 2.2, we have
$$\Sigma_{U,\tilde Z}
=\left(\begin{array}{ccc}0&(\Sigma_{U^*,U^*}-\Sigma_{U^*,Z}\Sigma_{Z,U^*})^{1/2}
\\Cov(\bar U^*,Z)
&Cov(\bar U^*,U^*)
\end{array}\right),$$ where $\bar U^*=(U^{( d+1)},\cdots,U^{(
p-q)})^\tau$. Furthermore,  from Lemmas~A1 and A2 in the Supplement, we can see that if
$$\gamma^\tau\Sigma_{U,\tilde Z}A^{\tau} (AA^{\tau})^{-1} A(\tilde
Z-E(\tilde Z))= \gamma^\tau\Sigma_{U,\tilde
Z}B^{\tau}(BB^{\tau})^{-1}B(\tilde Z-E(\tilde Z)),$$ model (\ref{2.3}) is
then unbiased, where $B=\left(\begin{array}{cccc} I_q & 0\\
A_1&A_2\end{array}\right)$ and $A=(A_1,A_2)$. Intuitively, when
$\|\gamma_{d+1}\mbox{corr}(U^{(d+1)}Z)\|_{\ell_1},
 \cdots,
\|\gamma_{p-q}\mbox{corr}(U^{(p-q)}Z)\|_{\ell_1}$ are small enough such that $Cov(\bar U^*,Z)$ is small enough,
the above equation can be approximately rewritten as
\begin{equation}
\label{4.1}\gamma^\tau\tilde\Sigma_{U,\tilde Z}A^{\tau} (AA^{\tau})^{-1} A(\tilde
Z-E(\tilde Z))= \gamma^\tau\tilde\Sigma_{U,\tilde
Z}B^{\tau}(BB^{\tau})^{-1}B(\tilde Z-E(\tilde Z)),\end{equation} where
$\tilde \Sigma_{U,\tilde Z}$ is an approximation to $
\Sigma_{U,\tilde Z}$ defined by
\begin{equation}
\label{4.2}\tilde \Sigma_{U,\tilde
Z}=\left(\begin{array}{ccc}0&(\Sigma_{U^*,U^*}-\Sigma_{U^*,Z}\Sigma_{Z,U^*})^{1/2}
\\0
&Cov(\bar U^*,U^*)
\end{array}\right).\end{equation} The matrix $\tilde\Sigma_{U,\tilde Z}$ has
rank $r=d$. Note that $d$ can be chosen to be small when the correlation
between $Z$ and $\bar U^*$ is weak and the inequality conditions in Proposition 2.1
hold. Consequently, the resultant $V$ is
a low-dimensional variable with dimension $r=d$. In some cases, $r$ can be $1$; for details see
Remark~4.1 below. By  the selected $\tilde\Sigma_{U,\tilde Z}$,
equation (\ref{4.1}) and the same arguments as in Sections~2,
we can get the corresponding orthogonal matrix $Q$ and a solution
$A=Q_1^\tau$ as given in Lemma~2.2. To guarantee the asymptotic
normality of the corresponding estimate of $\theta$, we need the
following condition on the extent of correlation between $Z$ and $U$:
\begin{itemize}
\item[\it C5.] Suppose $\|\mbox{corr}(U^{(d+1)}Z)\|_{\ell_1}
=O(n^{-\varsigma(d)})$, where $\varsigma(d)\geq 0$ depends on $d$.
The selected $d$ satisfies
$\varsigma(d)+\frac{1-c_1}{2}+\frac{m}{2m+d}>\frac12$, where
$0<c_1<1$ is determined in (\ref{2.1}) and $m$ is given in {\it
C2}.\end{itemize}  This condition is to rue out some cases in which removed predictors have too strong relationship with the kept predictors.  The asymptotic normality is stated below.

{\bf Theorem 4.1} {\it In addition to the conditions in Lemma~2.2,
assume that conditions C1-C4 and (\ref{2.2}) hold. When the selected $A$
is a solution of (\ref{4.1}), the selected $d$ satisfies C5 and $q$ is fixed, then, as
$n\rightarrow\infty$,
$$\sqrt{n}(\hat\theta-\theta)\stackrel{D}\longrightarrow
N(0,\sigma_V^2(Q_{12}^\tau Q_{12})^{-1}).$$

}


{\bf Remark 4.1} From Condition {\it C5}, we have the following
findings:
\begin{itemize}\item[(1)] If $0<c_1\leq 2\varsigma(d)<1$ for any $d$, Condition {\it C5} always holds for
any $d$. We then simply choose $d=1$. In this situation, the
selected quasi-instrumental variable $V$ is scalar and the
nonparametric estimation is only one-dimensional.
\item[(2)] For $d\in \mathscr{D}_1=\{d: 2\varsigma(d)< c_1<1\}$, we have $$
d\in \mathscr{D}_2=\left\{d:d<\frac{2m-2mc_1+4m\varsigma(d)}
{c_1-2\varsigma(d)}\right\}.$$ If $\mathscr{D}_1\cap \mathscr{D}_2$
is nonempty, the optimal choice of $d$ is
$$d^*=\min_{d}\{d:d\in\mathscr{D}_1\cap \mathscr{D}_2\}.$$
\end{itemize}

However, the above arguments only present some theoretical choices for $d$
since it depends on $c_1$ which describes the inexact-sparsity degree of
the model but usually  is unknown in advance. Without any prior
information about the inexact-sparsity, we suggest the following
selection approach.

{\bf A selection method for $d$.} Rank the values of
$\|\mbox{corr}(U^{(k)}Z)\|_{\ell_1},k=1,\cdots,p-q,$ in a descending
order $\|\mbox{corr}(U^{(1)}Z)\|_{\ell_1}\geq
\|\mbox{corr}(U^{(2)}Z)\|_{\ell_1}\geq \cdots \geq
\|\mbox{corr}(U^{(p-q)}Z)\|_{\ell_1}$. We first choose $d_1=1$ and
construct $U^*=U^{(1)}$. With the choices for $d$ and $U^*$ and the
method given in Subsection~2.2, we construct $\tilde Z$ and
calculate the rank $r_1$ of matrix $\Sigma_{U,\tilde Z}$. If
$r_1-d_1\leq 1$, we use $d=d_1=1$. If $r_1-d_1>
1$, we reselect $d_2=2$ and construct
$U^*=(U^{(1)},U^{(2)})^\tau$. Similarly, we further construct
$\tilde Z$ and calculate the rank $r_2$ of matrix $\Sigma_{U,\tilde
Z}$. Repeat this procedure until that $r_{m-1}-d_{m-1}> 1$ but
$r_m-d_m\leq 1$. Consequently, the final choice of $d$ is $d=d_m=m$.
The theoretical properties aforementioned ensure that such a choice,
$d=m$, is small under Condition~{\it C5}.

\subsection{Method 2}

If we are uncertain whether the selected $d$ satisfies Condition
{\it C5}, the above dimension-reduced method may not be efficient.
We now introduce an approximation to solve this problem. For the
convenience of representation, we here suppose $E(Z)=0$ and $
E(U)=0$. Recall that Lemma A3 given in the Supplement shows that
 model (\ref{2.3}) is unbiased if $A$ is a solution of the
following equation:
\begin{equation}
\label{4.3}\Sigma_{U,\tilde Z}{A}^{\tau}(A{A}^{\tau})^{-1}A\tilde Z
=\Sigma_{U,\tilde Z}B^{\tau}(BB^{\tau})^{-1}B\tilde Z.\end{equation} To
identify the quasi-instrumental variable $V$, we only need to
identify $A$. Thus, we suggest an approximation solution of (\ref{4.3})
with $A$ being supposed to be a row vector; in other words, we only
consider row vector solution of equation (\ref{4.3}). Of course this is
not an exact solution but  a low dimensional approximation.
Without confusion, we still use the notation $A$ to denote this row
vector.  For definitiveness of the choice of $A$, we constrain
$A{A}^{\tau}=1$. In this case $d=1$. From Lemmas A1 and A2, and the proof of Lemma~3.2 in the Supplement, 
we want to choose a row vector $A$ such that
\begin{equation}\label{4.4}{A}^{\tau}A\tilde Z=\Sigma_{U,\tilde Z}^+\Sigma_{U,\tilde Z}\tilde Z,\end{equation}
where $\Sigma_{U,\tilde Z}^+$ is the
Moore-Penrose generalized inverse matrix of $\Sigma_{U,\tilde Z}$.   By (\ref{4.4}), $A$ can be constructed as follows. Denote
$$A=(a_1,\cdots,a_q,a_{q+1}), \ A_k=a_kA, \ \Sigma_{U,\tilde
Z}^+\Sigma_{U,\tilde Z}=(D_1^{\tau},\cdots,
D_{q}^{\tau},D_{q+1}^{\tau})^{\tau},$$ where $D_k,k=1,\cdots,q+1$,
are $(q+1)$-dimensional row vectors. Note that $\Sigma_{U,\tilde
Z}^+\Sigma_{U,\tilde Z}=Q_1Q_1^\tau$. It can be consistently
estimated by the method proposed in the previous selection. Then we
estimate $A$ via solving the following optimization problem:
\begin{equation}
\label{4.5}\inf_{a_k,\cdots,a_{q+1}}\Big\{Q(a_1,\cdots,a_{q+1}):
\sum_{k=1}^{q+1}a_k^2=1\Big\},\end{equation} where
$Q(a_1,\cdots,a_{q+1})=\frac{1}{n}\sum_{i=1}^n\sum_{k=1}^{q+1}\|(A_k-D_k)\tilde
Z_i\|^2$. By the Lagrange multiplier, we obtain the estimates of
$A_k,k=1,\cdots,q+1,$ as
\begin{equation}
\label{4.6}\hat A_k=\Big(D_k\frac{1}{n}\sum_{i=1}^n\tilde Z_i {\tilde Z_i}^{\tau}+cc_ke_k/2\Big)\Big(\frac{1}{n}
\sum_{i=1}^n\tilde Z_i {\tilde  Z_i}^{\tau}+c_k
I\Big)^{-1},\end{equation} where $c_k>0$, which is similar to a ridge
parameter, depends on $n$ and tends to zero as $n\rightarrow\infty$,
and $e_k$ is the row vector with $k$-th component being 1 and the
others being zero. Note that the constraint $\|A\|=1$ implies
$\|A_k\|=\pm a_k$. By combining (\ref{4.6}) with these constraints we get
an estimate of $a_k$ as
$$\hat a_k=\pm \|\hat A_k\|$$ and consequently an estimate of $A$
is obtained by
$$\hat A=(\hat a_1,\cdots,\hat a_q,\hat a_{q+1}).$$
Finally, the estimated quasi-instrumental variable is
$$\hat V=\hat A\tilde Z.$$

\setcounter{equation}{0}
\section{ Simulation studies}

In this section we examine the performance of the new methods via
simulation studies. By mean squared error (MSE), model prediction
error (PE) and their $std$\,MSE and $std$\,PE as well, we compare
the new methods with the Gaussian-DS first. In ultra-high dimensional
scenarios, the DS cannot work well, we use the sure independent
screening (SIS) (Fan and Lv 2008) to bring the dimension $p$ down to a
moderate size and then to make a comparison with the Gaussian-DS.
As is well known, there are several factors that are of great impact
on the performance of variable selection methods: sparse or
non-sparse conditions, dimensions $p$ of predictor $X$, correlation
structure between the components of predictor $X$, and variation of
the error which can be measured by theoretical model R-square
defined by $R^2=(Var(Y)-\sigma^2_\varepsilon)/Var(Y)$. Then we will
comprehensively illustrate the theoretical conclusions and performances. {
In the present paper, we only focus on a comparison with the DS  because the initial working model is established by the DS in our paper. Clearly, this can also be applied to other methods such as the LASSO, the  SCAD or the MCP when the initial working models are selected by them. It deserves a further study.} 

{\bf Experiment 1.} This experiment is designed mainly for that with
different choices of the theoretical model R-square $R^2$, 
we compare our methods with
Gaussian-DS. 
In the simulation, to determine the regression coefficients, we
decompose the coefficient vector $\beta$ into two parts: $\beta_I$
and $\beta_{\bar I},$ where $I$ denotes the set of locations of
significant components of $\beta$. 
Three types of $\beta_I$ are considered:
\\ Type (I): $\beta_I= (1,0.4,0.3,0.5,0.3,0.3,0.3)^{\tau}$
and $I$= \{1,2,3,4,5,6,7\};\\ Type (II): $\beta_I=
(1,0.4,0.3,0.5,0.3,0.3,0.3)^{\tau}$ and $I= \{1,17,33,49,65,81,97\}$;\\
Type (III): $\beta_I= (1,0.4,-0.3,-0.5,0.3,0.3,-0.3)^{\tau}$ and $I=
\{1,2,3,4,5,6,7\}$. \\
To mimic practical scenarios, we set the values of the components
$\beta_{\bar Ii}$'s of $\beta_{\bar I}$ as follows. Before
performing the variable selection and estimation, we generate
$\beta_{\bar Ii}$'s from uniform distribution $\mathcal
{U}(-0.5,0.15)$ and the negative values of them are then set to be
zero. {
Thus the model under study here has around $23\%$ of coefficients in $\bar I$ are non-zero. In total, there are around 30 nonzero coefficients. Compared with the sample size $50$, this number is large.} After the
coefficient vector $\beta$ is determined, we consider it as a fixed
value vector and regard $\beta_I$ as the main part of the
coefficient vector $\beta$.
We use $\hat I$ to denote the estimate of $I$. Assume $X \thicksim
N_{p}(\mu,\Sigma_{X})$, where the components of $\mu$ corresponding
to $I$ are 0 and the others are 2, and the $(i,j)$-th element of
$\Sigma$ satisfies $\Sigma_{ij}=(-\rho)^{\mid i-j\mid}$, $0<\rho<1.$
Furthermore, the error term $\varepsilon$ is assumed to be normally
distributed as $\varepsilon \thicksim N(0,\sigma^2)$. In this
experiment, we choose different $\sigma$ to obtain different type of
full model with different $R^2$. In the simulation procedure, the
kernel function is chosen to be the Gaussian kernel
$K(u)=\frac{1}{\sqrt{2\pi}}\exp\{-\frac{u^2}{2}\}$, and bandwidths
for the nonparametric part of the estimation are chosen by the GCV
(generated cross-validation). The matrix $A$ is chosen respectively by Method 1 and
Method 2 suggested in Section~4. For Method 2, $A$ is determined by (4.6) with $c=2$,
$c_k=0.2$, $d=1$ and $U^{(1)}$ being the first component of $U$. The
choice of parameter $\lambda_p$ in the DS is just like that given by
Cand\'es and Tao (2007), which is the empirical maximum of
$|X^{\tau}z|_i$ over several
realizations of $z\sim N(0,I_n)$.

The following Tables 1 and 2 report the MSEs, the standard errors
(std) of MSE and the corresponding PEs via 200 repetitions with the sample size 50. In the two
tables, according to Methd 1 and Method 2, $\hat\beta^{(1)}_{\hat I}$ and $\hat\beta^{(2)}_{\hat I}$
are the corresponding estimates by (\ref{3.2}), $\tilde\beta_{\hat
I}$ is based on the Gaussian-dantzig selector. $\hat Y^{(1)}$ and $\hat Y^{(2)}$ are the corresponding predictions via the adjusted model (\ref{3.3}), $\hat Y_S^{(1)}$ and $\hat Y_S^{(2)}$ are in (\ref{3.4}) with the
new estimates $\hat\beta^{(1)}_{\hat I}$ and $\hat\beta^{(2)}_{\hat I}$,
$\tilde Y_S$ stands for the prediction in (\ref{3.5}).
The purpose of such a comparison is to see whether the
adjustment works, whether the working model (\ref{1.2}) should be used
when part of the components of high-dimensional data are not available
(say, too expensive to collect), and whether the new estimate
$\hat\beta_{\hat I}$ together with the bias-corrected model (\ref{2.3}) is
helpful for promoting prediction accuracy.

\newpage

\begin{center}
{ \small \centerline{{\bf Table~1.}     {\bf Simulation results for
Experiment 1 with $n=50,p=100$ and $\rho=0.1$}} \tabcolsep0.04in
\vspace{-1ex}
\newsavebox{\tablebox}
\begin{lrbox}{\tablebox}
\begin{tabular}{cc|ccc|ccccc|}
  \hline\hline
  & &\multicolumn{3}{c|}  {MSE($std$\,MSE)}  &\multicolumn{5}{c|}{PE($std$\,PE)} \\
 type&$R^2$&$\hat \beta^{(1)}_{\hat I}$ & $\hat \beta^{(2)}_{\hat I}$ &$\tilde\beta_{\hat I}$  &$\hat Y^{(1)}$& $\hat Y^{(1)}_S$ &$\hat Y^{(2)}$& $\hat Y^{(2)}_S$ & $\tilde Y_S$ \\\hline
       &0.97 &0.0076(0.0290) &0.0083(0.0458) &0.0661(1.0458)     &0.3446(0.0852) &0.3236(0.0834) &    0.3491(0.0867)  &  0.3484(0.0867)  & 1.8019(0.5057) \\
       &0.83 &  0.0155(0.0529)&  0.0178(0.0931)&   0.1837(0.2446)&0.7425(0.2062) &  0.7338(0.2004)  &        0.7478(0.2092) &   0.7455(0.2098) &   1.9524(0.4851)     \\
(I)    &0.68 &0.0280(0.0915) & 0.0322(0.1581) &  0.3111(0.3314)&1.3007(0.3370)  &  1.2728(0.3380) &     1.3886(0.3844) &   1.3831(0.3848) &            3.1355(0.6860)  \\
       &0.51 & 0.0554(0.2166)&  0.0528( 0.2166)&  0.3700(0.2166)&2.5164(0.6858)  & 2.0656(0.7858) &    2.4629(0.5441)  &  2.4486(0.5334)    &     3.2328(0.7167)       \\
       &0.34 &   0.1141(0.6882)& 0.1098(0.5994)&   0.4231(0.5719)&5.6583(1.0436)&   4.6131(1.0436) & 5.5521(1.4106)& 5.4727(1.3625)  &  5.6865(1.2511)       \\
\hline
      &0.98 &0.0057(0.0246) & 0.0062(0.0393)&  0.2666(0.0393)&0.2523(0.0717)  &  0.2544(0.0624) &  0.2646(0.0709)  &  0.2635(0.0705) &   2.7789(0.7396) \\
      &0.82 & 0.0176(0.0727)&  0.0202(0.1280)&  0.3360(0.3658)&0.9268(0.2431)&    0.8748(0.2208)& 0.8432(0.3032)& 0.8431(0.3057)& 3.2786(0.7576) \\
(II)  &0.70 & 0.0371(0.1680) & 0.0410(0.2412) & 0.0821(0.2060)& 1.8231(0.6182) &   1.8211(0.6223)  & 1.8353(0.4151)  &  1.8258(0.4169)  &  2.2044(0.4473) \\
      &0.53 & 0.0535(0.2347) & 0.0590(0.3274)&  0.1346(0.3146)&2.4823(0.8481)&   2.4203(0.8581) & 2.5878(0.8385)& 2.5820(0.8333) & 3.3047(0.7489) \\
      &0.33 &  0.1075(0.5257) &0.1127(0.6070) &  0.1380(0.5562)&5.4846(1.2681)& 5.6176(1.2625) & 5.4149(1.3041) & 5.3628(1.2690) & 5.6352(1.2625) \\
\hline
      &0.98 &0.0066(0.0247)&  0.0066(0.0256)&  0.1745(0.5922)&0.2890(0.0721) &   0.2728(0.0709) &    0.2934(0.0777)&    0.2908(0.0760)& 8.5007(1.5584) \\
      &0.83 & 0.0207(0.0960) & 0.0218(0.0982) & 0.2916(0.3477)&1.1471(0.3021)  &  0.8076(0.3117)&     0.9325(0.2566) &   0.9277(0.2574)   &  3.0455(0.7593) \\
(III) &0.67 &0.0359(0.1738)& 0.0330(0.1974) &  0.1253(0.2713)&1.8805(0.6640)&   1.2958(0.5864)&    1.6082(0.5810)&    1.5998(0.5777) & 2.3560(0.5576)\\
      &0.49 &  0.0538(0.2525)& 0.0587(0.2527) &  0.0926(0.2454)&2.7107(0.6603) & 2.6685(0.6629) & 2.6299(0.6564) &2.6178(0.6584) & 2.9190(0.7505) \\
      &0.30 &0.1111(0.8190)  &  0.1173(0.6136)  &  0.1117(0.5611)&5.5375(1.4581) &   5.5381(1.4574) &     5.5532(1.5246) &   5.5018(1.5131)& 5.6872(1.2333)          \\
\hline\hline
\end{tabular}
\end{lrbox}
\scalebox{0.7}{\usebox{\tablebox}} }
\end{center}

\

\begin{center}
{ \small \centerline{{\bf Table 2.}     {\bf Simulation results for
Experiment 1 with $n=50,p=100$ and $\rho=0.7$}} \tabcolsep0.04in\vspace{-1ex}
\begin{lrbox}{\tablebox}
\begin{tabular}{cc|ccc|ccccc|}
  \hline\hline
  & &\multicolumn{3}{c|}  {MSE($std$\,MSE)}  &\multicolumn{5}{c|}{PE($std$\,PE)} \\
 type&$R^2$&$\hat \beta^{(1)}_{\hat I}$ & $\hat \beta^{(2)}_{\hat I}$ &$\tilde\beta_{\hat I}$  &$\hat Y^{(1)}$& $\hat Y^{(1)}_S$ &$\hat Y^{(2)}$& $\hat Y^{(2)}_S$ & $\tilde Y_S$ \\\hline
&0.96 & 0.0212(0.0654) & 0.0210(0.0627) & 0.3253(0.3440)&0.3817(0.0964)& 0.3591(0.1155) &0.3873(0.0978) &   0.3802(0.0911) &   1.9617( 0.4463) \\
& 0.72 &    0.0426(0.1029)& 0.0426(0.1360) &  0.7535(0.3552)&0.8480(0.1762)    &0.7899(0.1759) &0.7208(0.1760) &0.7057(0.1651) &3.1180(0.6025)
 \\
(I)&       0.53 &   0.0413(0.1368)&0.0449(0.1602)&  0.2704(0.2870)&1.3493(0.3340) &   1.1855(0.4340) & 1.2676(0.3513)& 1.2424(0.3279)&2.4560(0.6153)
 \\
&       0.33 &    0.0710(0.2402)&0.0796(0.2982) & 0.1532(0.2356)&2.3931(0.6414) &1.5336( 0.6561)&2.2808(0.6312)& 2.2595(0.6313)& 2.6386(0.6261)
 \\
&        0.2 &    0.1506(0.9474)&0.1550(1.0552)&0.1602(0.8273)&5.2751(4.6510) &5.2837(1.3568)& 5.2913(1.5935)&5.2669(1.5729)&5.3565(1.4338)
 \\\hline
&       0.98 &    0.0138(0.0683)& 0.0134(0.0489) & 0.2160(0.7071)&0.4884(0.2773) & 0.3923(0.1149)& 0.4211(0.1153) &   0.4192(0.1147) &  10.1951(1.8591) \\
&       0.84 &    0.0228(0.0910)&0.0195(0.1919)&0.1679(0.1977)&0.9586(0.1982)&0.7125(0.2530)& 0.8172(0.1837)&0.8142(0.1822) &1.6597(0.3255)\\
(II)&       0.69 &0.0445(0.1632)& 0.0506(0.3019)& 0.2099(0.2525)& 1.9009(0.5047) &   1.3173(0.5425)&1.7545(0.4501)&1.7358(0.4444)&2.4400(0.5553)\\
&       0.52 &   0.0619(0.2690)&0.0445(0.1632)&0.2100(0.3414)&2.1641(1.4330)&2.0742(0.9001)&1.9009(1.4047)&1.3173(0.6425)& 3.2587(0.8246)\\
&       0.35 &    0.1190(0.7436)& 0.1254(0.8459)&0.1318(0.6574)&6.0371(1.4171)&3.8815(2.1353)&5.3514(1.4601)&5.3077(1.4338)&5.3859(1.3776)
 \\\hline
&       0.96 &     0.0222(0.0605)&  0.0215(0.0517) & 0.2365(0.8033)& 0.2907(0.0631)&0.2440(0.0576) &  0.2852(0.1162)&0.2572(0.0694)&6.4491(0.7604)
\\

           &       0.74 &    0.0382(0.1193)&0.0415(0.1189)&0.5241(0.4315)&0.9856(0.2315)& 0.8237(0.2365)&0.8912(0.2612)&0.8366(0.2039)&3.2644(0.7906) \\

           (III)&       0.56 &0.0539(0.1458)& 0.0564(0.2124)&0.1793(0.2335)& 1.2743(0.3813)&1.0626(0.3731)&1.2528(0.3534)&1.2182(0.3171)&1.8547(0.4177)

\\

        &       0.39   &      0.0748(0.2815) & 0.0679(0.3149) & 0.2179( 0.3525)&2.2543(1.7217)&1.5985(1.1337) &2.4484(0.6324)&2.4401(0.6290)&3.3653(0.8554)
 \\

           &       0.23 &    0.1308(0.6035)& 0.1206(0.4495)&0.1351(0.3674)&4.7107(1.5897)&4.5846(1.8813)&4.5947(1.3848)&4.4944(1.2940)&4.7708(1.0775)
 \\

\hline\hline
\end{tabular}
\end{lrbox}
\scalebox{0.7}{\usebox{\tablebox}} }
\end{center}

\

The simulation results in Table 1 suggest that  the adjustments of
(\ref{3.3}) and (\ref{3.4}) work very well in the sense that the corresponding estimates $\hat
\beta^{(k)}_{\hat I}$ and predictions $\hat Y^{(k)}$ and $\hat Y^{(k)}_S$ ($k=1, 2 $) are uniformly best than the competitor $\tilde Y_S$, and the differences between $\hat Y^{(k)}$ and $\hat Y^{(k)}_S$ ($k=1, 2 $) are small.
To provide more information, we also consider the case with higher
correlation between the components of $X$. Table~2 shows that when
$\rho$ is larger, the conclusions from  the
comparison are almost identical to those obtained from Table~1. 
Thus from the limited simulations we conduct, it concludes that no matter $\rho$ is larger or not, for
different choices of $R^2$, our  method  works  well.

We are now in the position to make another comparison. In
Experiments~2 and 3 below, we do not use the data-driven approach as
given in Experiment 1 to select $\lambda_p$, while manually select
several values to see whether our method works or not. This is
because in the two experiments, it is not our goal to study
shrinkage tuning parameter, but is our goal to see whether the new
method works after we have a working model no matter the working model is a ``right model" including all ``significant predictors" or not.

{\bf Experiment 2.}  In this experiment, our focus is that with
different choices of working
models and the correlation between predictors, we compare our method with the others. The distribution of $X$
is the same as that in Experiment~1 except for
different dimension of $X$. 
The coefficient vector $\beta_I$ is designed as type (I) above and
$\beta_{\bar I}$ is designed as in Experiment 1. Thus the model here
is also non-sparse.
 Furthermore, the error term
$\varepsilon$ is assumed to be normally distributed as $\varepsilon
\thicksim N(0,0.2^2)$.  

As different choices of $\lambda_p$ usually lead to different
working models, equivalently, to different estimates $\hat I$ of
$I$, we then consider different choices of $\lambda_p$ in the
simulation
study. 
The setting is as follows. For $n=50,p=100$ and
$\rho=0.1,0.3,0.5,0.7,$ we consider two cases
for each $\rho$: \\
$\rho=0.1:\\
$ Case 1. $\lambda_p=3.97,$ $I$=\{1,2,3,4,5,6,7\}, $\hat I$=\{1,     2,     3,     4,     5,     6,     7 \}\\
Case 2. $\lambda_p=6.53,$ $I$=\{1,2,3,4,5,6,7\}, $\hat I$=\{ 1,  2,  3,    4,   25 \}\\
$\rho=0.3:\\$ Case 1. $\lambda_p=3.32,$ $I$=\{1,2,3,4,5,6,7\}, $\hat I$=\{ 1,     2,   4   ,  5,  6,  7,   20 \}\\
Case 2. $\lambda_p=6.77,$ $I$=\{1,2,3,4,5,6,7\}, $\hat I$=\{   1,   2,    4,    7,  24 \}\\
$\rho=0.5:\\$ Case 1. $\lambda_p=3.72,$ $I$=\{1,2,3,4,5,6,7\}, $\hat I$=\{ 1,    2,     4,     6,    7\}\\
Case 2. $\lambda_p=7.29,$ $I$=\{1,2,3,4,5,6,7\}, $\hat I$=\{ 1,     2,    4 \}\\
$\rho=0.7:\\$ Case 1. $\lambda_p=3.50,$ $I$=\{1,2,3,4,5,6,7\}, $\hat I$=\{1,   2,   3,     4\}\\
Case 2. $\lambda_p=7.22,$ $I$=\{1,2,3,4,5,6,7\}, $\hat I$=\{ 1, 4,  43,   99 \}\\

\begin{center}
{ \small \centerline{{\bf Table 3.}     {\bf Simulation results for
Experiment 2 with $n=50,p=100,S=7$} } \tabcolsep0.045in\vspace{-1ex}
\begin{lrbox}{\tablebox}
\begin{tabular}{cc|ccc|ccccc|}
  \hline\hline
  & &\multicolumn{3}{c|}  {MSE($std$\,MSE)}  &\multicolumn{5}{c|}{PE($std$\,PE)} \\
 $\rho$&Case&$\hat \beta^{(1)}_{\hat I}$ & $\hat \beta^{(2)}_{\hat I}$ &$\tilde\beta_{\hat I}$  &$\hat Y^{(1)}$& $\hat Y^{(1)}_S$ &$\hat Y^{(2)}$& $\hat Y^{(2)}_S$ & $\tilde Y_S$ \\\hline
%
                          &1&   0.0078(0.0434)&  0.0070(0.0322)&   0.3804(1.4008)& 0.2556(0.2021 )&    0.3005(0.0853) &       0.3195(0.0861)   & 0.3187(0.0859)  &   17.6573(2.6412)
\\
\raisebox{1.5ex}[0pt]{0.1}&2 &  0.0122(0.0464)& 0.0120(0.0120)  & 0.4385(0.4904)&0.5457(0.2353)&    0.5379(0.1656)&      0.5179(0.1460) &  0.5138(0.1428)&3.1373(0.8837)
\\\hline
                          &1 &    0.0085(0.0432)& 0.0083(0.0364)&   0.1679(0.2022)& 0.3531(0.1531) &  0.2410(0.0832)&        0.2685(0.0638) &   0.2660(0.0635) &     1.6592 (0.3298)

\\
\raisebox{1.5ex}[0pt]{0.3}&2 &    0.0164(0.0671)& 0.0145(0.0495) &  0.2624(0.2871)&0.4524(0.1784)& 0.4468(0.1346) & 0.4033(0.1026) & 0.4008(0.1020) &1.9283(0.4689)
\\\hline
                          &1 &   0.0169(0.0400) &0.0169(0.0420) &  0.1099(0.4070)& 0.3738(0.1059)&   0.2328(0.0637)&         0.2626(0.0730)&    0.2376 (0.0517) &  3.9955(0.5713) \\
\raisebox{1.5ex}[0pt]{0.5}&2 &   0.0213(0.0421) &0.0220(0.0561)  & 0.2448(0.6498)& 0.5698(0.1132) &  0.3591(0.0963)&         0.4178(0.1047)&    0.3856(0.0885) &    7.4154(0.9678)
\\\hline
                          &1 &    0.0205(0.0720)&  0.0199(0.0874) &  0.4473(1.5829)& 0.2829(0.0907)&    0.2815(0.0695)  &     0.2981(0.1057)&0.2822(0.0963)& 8.2681(1.0685)
\\
\raisebox{1.5ex}[0pt]{0.7}&2 & 0.0194(0.0450)  &0.0200(0.0534) &  0.1987(0.1364)& 0.2986(0.1329) &   0.2365(0.0761) &   0.2699(0.0676) &   0.2618(0.0651)  &  1.0810(0.2836)
\\
\hline\hline
 \end{tabular}

\end{lrbox}
 \scalebox{0.7}{\usebox{\tablebox}} }
\end{center}

\

From Table~3, we can see clearly that the correlation is of impact
on the performance of the variable selection methods:  the
estimation gets worse with larger $\rho$. However, the new method
uniformly works much better than the Gaussian DS, when we compare
the performance of the methods with different values of $\lambda_p$
and then with different working models. We can see that in Case I,
the working models are more accurate than those in Case II in the
sense that they can contain more significant predictors we want to
select. Then, the estimation based on the Gaussian Dantzig
selector can work better and so can the new method. 

In the following, we consider data with higher-dimension.

{\bf Experiment 3.} In this experiment $\beta_{\bar I}$ is designed
as in Experiment 1. When $p=1000$, there are around 230 nonzero coefficients. This number is much larger than the sample size $100$ in this experiment.  Thus the model here is also non-sparse. For very
large $p$, the DS method alone cannot work well. Thus, we use the
sure independent screening (SIS,  Fan and Lv 2008) to reduce the
number of predictors to a moderate scale smaller than the sample
size, and perform  variable selection and parameter
estimation afterwards by the
Gaussian DS and our adjustment method. {
Therefore, almost all coefficients after the sure screening are non-zero. This is a very typical non-sparse model with bias.} 
The experiment conditions are designed as: $$\beta_I=(1.0, -1.5,
2.0, 1.1, -3.0, 1.2, 1.8, -2.5, -2.0, 1.0)^{\tau},n=100,p=1000;$$
\noindent $\rho$=0.1:  \\
Case 1. $\lambda_p$=4.50,  $I$=\{1,2,3,4,5,6,7,8,9,10\}, $\hat I=\{1,     2,     3,     4,     5,     7,     8,     9
\}$;\\
Case 2. $\lambda_p$=7.30,  $I$=\{1,2,3,4,5,6,7,8,9,10\}, $\hat I$=\{ 3,     5,     8,     9\};\\
$\rho$=0.5: \\
Case 1. $\lambda_p$=3.56,  $I$=\{1,2,3,4,5,6,7,8,9,10\}, $\hat I=\{1,     2,     5,     6,     7,     8,     9\}$;\\
Case 2. $\lambda_p$=6.92,  $I$=\{1,2,3,4,5,6,7,8,9,10\}, $\hat I=\{2,     5,     8,     9 \}$;\\
$\rho$=0.9:\\
Case 1. $\lambda_p$=1.80,  $I$=\{1,2,3,4,5,6,7,8,9,10\}, $\hat I=\{  1,     2,     5,     9
 \}$;\\
Case 2. $\lambda_p$=5.83,  $I$=\{1,2,3,4,5,6,7,8,9,10\}, $\hat I=\{
1,     5 \}.$

With this design, $\lambda_p$ in Case I results in that  more
significant predictors are selected into the working model than
those in Case II so that we can see the advantage of the adjustment
method.

\newpage

\begin{center}
{ \small \centerline{{\bf Table 4.}     {\bf Simulation results for
Experiment 3 with $n=100$ and $p=1000$ }} \tabcolsep0.04in
\vspace{-1ex}
\begin{lrbox}{\tablebox}
\begin{tabular}{cc|ccc|ccccc|}
  \hline\hline
  & &\multicolumn{3}{c|}  {MSE($std$\,MSE)}  &\multicolumn{5}{c|}{PE($std$\,PE)} \\
 $\rho$&Case&$\hat \beta^1_{\hat I}$ & $\hat \beta^2_{\hat I}$ &$\tilde\beta_{\hat I}$  &$\hat Y^1$& $\hat Y^1_S$ &$\hat Y^2$& $\hat Y^2_S$ & $\tilde Y_S$ \\\hline
                          &1&  0.7354(0.5297) &    0.5802(0.4485) &  30.2566(18.3643)&6.9334(2.1769) &  6.9262( 1.9136)&    5.6459(1.6411)&    5.5941(1.6289) & 357.5061(81.1949)
 \\
\raisebox{1.5ex}[0pt]{0.1}&2&   1.2328( 1.9193) &1.0982(1.7057)&13.9411(11.0267)& 12.5241(6.3322)   & 12.1051(6.9150)  & 13.7420(5.0803)&   13.5899(5.0010)&  363.9597(75.1476)
 \\\hline
                          &1& 2.7609(2.6864)  &  2.2720(1.2278)&   57.4401(36.1276)& 14.2629(3.9249)&   17.7230(4.5973)&     12.1683(3.1217)   &10.7383(2.3452) & 513.6982(69.5787)
\\
\raisebox{1.5ex}[0pt]{0.5}&2&6.0680(1.6366)  &  5.8816(1.9266) &  25.0097(17.1664)& 16.1536(5.4010) &   17.9348( 5.2269) &   16.8991(4.2717) &   15.5496(3.8825) &  347.1439(44.3003)
\\\hline
                          &1&  10.0895(5.7459) &   9.6210(4.1308) &  69.3674(65.3793)& 15.2553(4.5103) &  12.5484(5.1151 )  &  12.3265(4.3022)  & 11.2805(3.1104) & 349.6882(44.2636)\\
\raisebox{1.5ex}[0pt]{0.9}&2&2.6881(1.6762)&    2.5246(1.7219)&   16.0189(18.2411)&16.2310(3.4421) &  16.5675(3.3055)&   13.8226(3.1822)&   13.2988(2.8200)&  370.0862(35.8113)
\\
\hline\hline
 \end{tabular}

\end{lrbox}
\scalebox{0.7}{\usebox{\tablebox}} }
\end{center}

\

From Table 4, we have the conclusion that the SIS does work to
reduce the dimension such that the Gaussian DS and our method can be
efficiently performed. In the scenarios with small and large  correlation coefficients
$\rho$ being from 0.1 to 0.9, the new method works
better than the Gaussian DS. The conclusions are almost identical to
those with smaller $p$ in Experiments~1 and 2. Thus, we do
not give more comments here. Further, by comparing the results of
Cases I and II, we can see that the adjustment can work better
when the working model is not well
selected. 

{\bf Experiment 4.} This experiment is used to examine the performance of our method for sparse models.  We also consider three types of $\beta$ which
are the same as those in Experiment~1 except that all the components of
$\beta_{\bar I}$ are zero. The simulation results are reported in
Table~5 below.

\newpage

\begin{center}
{ \small \centerline{{\bf Table 5.}     {\bf Simulation results for
Experiment 4 with sparsity condition and $n=50$, $p=100$ and $\rho=0.3$}}
\tabcolsep0.04in \vspace{-1ex}
\begin{lrbox}{\tablebox}
\begin{tabular}{cc|ccc|ccccc|}
  \hline\hline
  & &\multicolumn{3}{c|}  {MSE($std$\,MSE)}  &\multicolumn{5}{c|}{PE($std$\,PE)} \\
 type&$R^2$&$\hat \beta^1_{\hat I}$ & $\hat \beta^2_{\hat I}$ &$\tilde\beta_{\hat I}$  &$\hat Y^1$& $\hat Y^1_S$ &$\hat Y^2$& $\hat Y^2_S$ & $\tilde Y_S$ \\\hline

   & 0.97 &    0.0011(0.0056)&  0.0011(0.0057)  & 0.0009( 0.0040)&0.0501(0.0129)&     0.0470(0.0110)  &     0.0480(0.0134)&     0.0479( 0.0133)  &       0.0466(0.0094)
 \\
(I)&0.81 &       0.0161(0.0816)& 0.0152(0.0630) &  0.0120(0.0366)&
   0.6252( 0.1556)   & 0.6475( 0.1979)&         0.6154(0.1692)&    0.6126(0.1686)&    0.5868(0.1609) \\
   &       0.66&    0.0318(0.1522)&  0.0336(0.1568)  &0.0231(0.0914)&1.5207(0.4457) &   1.5851(0.5271)&         1.4396(0.3857) &   1.4344(0.3850) &   1.3817(0.3489)

 \\

   &        0.48 &   0.0589(0.2842) & 0.0571(0.2849) & 0.0392(0.1529)& 2.5454(0.8854) &  2.5617(0.9552)   &     2.4978(0.6834)  &  2.4859  (0.6812)  &   2.3777(0.5940)
\\
   &0.30 &0.1253(0.5924)& 0.1155(0.5753)&   0.0950(0.4254)&5.3103(1.6553)&    5.2178(1.6782 ) &   5.3244(1.2781) &   5.2417( 1.1567)& 5.3643(1.1403)           \\\hline

    &       0.98 &    0.0011(0.0053) & 0.0011(0.0055)&  0.0010(0.0040)&0.0512(0.0121)   & 0.0461(0.0133)   &      0.0485(0.0117)&    0.0484(0.0116)  &  0.0467(0.0111)
 \\

    &       0.83 &   0.0147(0.0669) &0.0143(0.0632)&   0.0123(0.0400)& 0.6233(0.2892) &   0.5885(0.2228)  &   0.6131(0.1922) &   0.6113(0.1924) &       0.5914(0.1389)
 \\

(II)&0.69&       0.0204(0.0967)&  0.0184(0.0686) &  0.0141(0.0538)& 0.8922(0.2691) &    0.7506(0.2918)         &  0.8625(0.1963) &   0.8559 (0.1858)&     0.8214(0.1710)
 \\
    &       0.51 &     0.0607(0.4194) &  0.0632(0.3031)&  0.0376(0.1320)& 2.6871(0.8511) &    1.9209(0.8912)  &       2.4965(0.7490)&     2.4831(0.7422) &      2.3767(0.5534)
 \\

    &       0.32 &   0.1157(0.6849)&  0.1179(0.6243)&   0.0920(0.5110)&
   5.5929(1.4590) &    4.9211(1.7027)  &        5.4506(1.4685) &   5.4462(1.4337) &             5.2151(1.2803)

\\\hline

           &       0.98 &    0.9803$\times 10^{-3}$(0.0049)  &  0.0012(0.0060)  &  0.0011(0.0044)&0.0600(0.0342) &  0.0452(0.0136)     &   0.0508(0.0130) &   0.0505(0.0128) &   0.0480(0.0113)
 \\

           &       0.81 &     0.0161(0.0816)& 0.0152( 0.0630) &  0.0120(0.0366)&0.7352(0.2556) &   0.5775(0.1979)    &        0.6154(0.1692)  &    0.6126(0.1686)  &    0.5868(0.1609)
 \\

    (III)&       0.66 &   0.0318(0.1522) & 0.0336(0.1568)&  0.0231(0.0914)&1.6207(0.3457)  &  1.0851(0.5271)   &       1.4396(0.3857) &     1.4344(0.3850) &     1.3817( 0.3489)

 \\

           &       0.48 &    0.0589(0.2842) &  0.0571(0.2849)& 0.0392(0.1529)&2.5454(0.7854)  &  2.5617(0.9552)   &     2.4978(0.6834)  &  2.4859(0.6812) &    2.3777(0.5940)

 \\

           &       0.30 &   0.1253(0.5924)  &0.1155(0.5753)  &  0.0950(0.4254) &5.6103(1.1553)&    5.0178(1.0782)    &   5.3244(1.2781)  &  5.2417(1.1567)&      5.0643(1.1403)
 \\

\hline\hline
\end{tabular}
\end{lrbox}
\scalebox{0.7}{\usebox{\tablebox}} }
\end{center}

\

From this table, we can see that even in sparse cases, for every
type of $\beta$, the new estimate $\hat \beta^{(k)}_{\hat I},k=1, 2,$ are in
almost all cases better than $\tilde\beta_{\hat I}$  in the
sense of smaller MSE. This is also the case for prediction: $\hat Y^{(k)}, k=1, 2,$
have smaller prediction error than $\tilde Y_S$ does when $\rho\ge
0.1$. It is not a surprise that $\hat Y^{(k)}_S, k=1, 2,$ cannot be as good as its
performance in non-sparse cases, but still comparable to $\tilde
Y_S$.
Overall, the new method is still
comparable to the classical method in the sparse models under study.

In summary, the results in the six tables above obviously show the
superiority of the new estimate $\hat\theta$ and the bias-corrected
model (\ref{2.3})/the working model (\ref{1.2}) over the others in the literature
with smaller MSEs, PEs and standard errors.
The good performance holds for
different combinations of the sizes of selected working models
(values of $\lambda_p$), $n,p,S,I$, $R^2$ and the correlation
between the components of $X$. The new method is particularly useful
when a submodel, as a working model, is very different from
underlying true model. Thus,  the adjustment method is  worthy of
recommendation. Also it is comparable to the classical method in
sparse cases, suggesting its robustness against model structure.
However, as a trade-off, the adjustment method involves
nonparametric estimation, although low-dimensional ones. It makes
estimation not as simple as that obtained by  existing ones.
Thus, we may consider using it after a check whether the submodel is
significantly biased. The relevant research is ongoing.

\

\noindent{\large\bf  Supplementary Materials.}
\begin{description}

     \item[Proofs of the theorems:] The pdf file  ``supplement-4.pdf" containing detailed proofs of the lemmas and theorems.
     \item[Matlab package for DANTZIG CODE routine:] Matlab package "DANTZIG CODE"  containing the codes. (WinRAR file)

\end{description}


\bibliography{dantzig_two_stage}
\

\leftline{\large\bf References}
\begin{description}

\baselineskip=15pt






%

%

\item Bickel, P. J., Ritov, Y. and Tsybakov, A. (2009). Simultaneous
analysis of LASSO and Dantzig selector.  {\it Ann. Statist.}, {\bf
37}, 1705-1732.

\item Cand\'es, E. J. and Tao, T. (2007). The Dantzig selector:
statistical estimation when $p$ is much larger than $n$. {\it Ann.
Statist.} {\bf 35}, 2313-2351.





\item Fan, J. and Li, R. (2001). Variable selection via nonconcave
penalized likelihood and its oracle properties. {\it J. Am. Statist.
Ass.}, {\bf 96}.

\item Fan, J. and Peng, H. (2004). Nonconcave penalized likelihood
with a diverging number of parameters. {\it Ann. Statist.}, {\bf
32}, 928-961.

\item Fan, J., Peng, H. and Huang, T. (2005). Semilinear
high-dimensional model for normalized of microarray data: a
theoretical analysis and partial consistency. {\it J. Am. Statist.
Ass.}, (with discussion), {\bf 100}, 781-813.

\item Fan, J. and Lv, J. (2008). Sure independence screening for
ultrahigh dimensional feature space. {\it J. R. Statist. Soc.} B
{\bf 70}, 849-911.



\item H\"ardle, W., Liang, H. and Gao, T. (2000). {\it Partially linear
models.} Physica Verlag.

\item
Hall, P. and Li, K. C. (1993). On almost linearity of low
dimensional projection from high dimensional data. {\it Ann.
Statist.} {\bf 21}, 867-889.

\item
Leeb, H. and P\"otscher, B. M. (2005). Model selection and
inference: facts and fiction. {\it Econometric Theory}, {\bf 21},
21-59.





\item Lin, L., Zhu, L. X. and Gai, Y. J. (2010). Adaptive post-Dantzig estimation and prediction for non-sparse `` large $p$ and small $n$" models.  {\it arXiv:1008.1345v1 Stat.ME 7 Aug 2010}, 1 -- 37.


%

%





\item Tibshirani, R. (1996). Regression shrinkage and selection via the lasso. {\it J. Royal. Statist. Soc B.}, {\bf 58}, 267 - 288.
\item Zhang, C. H. (2010),
Nearly unbiased variable selection under minimax concave penalty.
\textit{The Annals of statistics}, {\bf 38}, 894--942.

\item Zhang, C. and Huang, J. (2008). The sparsity and bias of the
LASSO selection in high-dimensional linear regression. {\it Ann.
Statist.}, {\bf 36}, 1567-1594.



\item Zou, H. (2006). The adaptive lasso and its oracle properties. {\it J. Am. Statist.
Assoc.}, {\bf 101}, 1418-1429.

\end{description}

\end{document}